\documentclass[12pt]{article}

\usepackage{amssymb}
\usepackage{amsmath}

\usepackage{graphicx}

\begin{document} %


\title{A search for heavy axion-like particles in light-by-light scattering at the FCC-hh}

\author{
S.C. \.{I}nan\thanks{Electronic address: sceminan@cumhuriyet.tr}
\\
{\small Department of Physics, Sivas Cumhuriyet University, 58140,
Sivas, Turkey}
\\
{\small and}
\\
A.V. Kisselev\thanks{Electronic address:
alexandre.kisselev@ihep.ru} \\
{\small A.A. Logunov Institute for High Energy Physics, NRC
``Kurchatov Institute''},
\\
{142281, Protvino, Russian Federation}}

\date{}

\maketitle

\begin{abstract}
A virtual production of heavy axion-like particles (ALPs) via
light-by-light scattering in pp, pPb and Pb collisions at the future
100 TeV collider FCC-hh is studied. Both differential and total
cross sections are calculated. The 95\% C.L. exclusion limits, as
well as $3\sigma$ and $5\sigma$ discovery limits on an ALP coupling
constant versus ALP mass $m_a$ are given, using integrated
luminosities of 30 ab$^{-1}$, 27 pb$^{-1}$ and 110 nb$^{-1}$. Our
results are compared with the current LHC bounds. The strongest
limit on the ALP coupling is obtained if $m_a \simeq 250$ GeV for
the PbPb collisions, and if $m_a \simeq 1$ TeV for pp or pPb
collisions. This suggests that the FCC-hh has a great
physics potential of searching for the heavy ALPs.
\end{abstract}

\maketitle


\section{Introduction} %

In our previous papers we examined a production of axion-like
particles (ALPs) in a number of processes at the CLIC
\cite{I-K:2020}-\cite{I-K:2025} and future muon collider
\cite{I-K:2023}. The notions of axion and ALP arise in the framework of the
so-called strong CP problem. This problem is one of the open issues of
the Standard Model (SM). It can be solved by introducing a
spontaneously broken global Peccei-Quinn (PQ) symmetry
\cite{Peccei:1977_1,Peccei:1977_2}. This results in a light
pseudoscalar Nambu-Goldstone boson called QCD axion
\cite{Weiberg:1978,Wilzcek:1978}. There are two main models in which
the PQ symmetry is decoupled from the electroweak scale and is
spontaneously broken \cite{Kim:1979}-\cite{Zhitnitsky:1980} (for a
systematic compilation of theoretical axion-like models, see recent
paper \cite{WISPedia}). For the time being, the very low-mass and
weakly coupled axion is a well motivated candidate for the dark
matter \cite{Preskill:1983}-\cite{Lentz:2026}.

In addition to the QCD axion, a broader class of ALPs appears in
many theories beyond the SM, such as string-inspired
scenarios, grand unified theories, and models which include
spontaneously broken global symmetries. Unlike the QCD axion,ALPs are characterized by independent mass and coupling parameters, which makes
them a powerful and flexible framework for searches beyond the
Standard Model.

The strong interest in axions has extended to ALPs. It is
assumed that the ALP is a particle having interactions similar to
the axion, but without any relationship between its coupling to the
SM fields $f_a$ and mass $m_a$. This implies that $f_a$ and $m_a$ can be
treated independently. The ALPs emerge in string theories, in models
with spontaneously broken symmetries, and in GUT scenarios. All
these models predict an ALP-photon coupling and, therefore, an
electromagnetic production of the ALPs and their decay in two
photons.

This coupling to photons makes photon-induced processes at
high-energy colliders particularly appropriate for ALP searches.
Light-by-light (LbL) scattering also plays a special role as it is
a rare SM scattering occurring at loop level. Therefore, it is highly
sensitive to new physics contributions.

The heavy ALPs can be detected at present and future colliders
\cite{Mimasu:2015}-\cite{Biekotter:2025}. The production of the ALPs
in LbL scattering at the LHC was experimentally studied both in
proton-proton collisions
\cite{ATLAS_pp_axion:2014}-\cite{CMS_pp_axion:2024} and lead-lead
collisions \cite{CMS_LbL_axion:2019}-\cite{CMS_LbL_axion:2025}. In a
large number of papers \cite{Beldenegro:2017}-\cite{Barbosa:2025} a
phenomenological analysis of the ALPs in pp and heavy ion processes
at the LHC was presented. Studies of ALP searches at
future colliders can be found in \cite{Feng:2025},
\cite{Coelho:2020_1}-\cite{Coelho:2020_3}, and
\cite{Bauer:2017_2}-\cite{Adhikary:2026}.

In this paper, we study an exclusive two-photon
production mediated by ALPs in the following collisions at the
Future Circular Collider in hadron-hadron mode (FCC-hh)
\cite{FCC:V_3}-\cite{FCC:physics}:
\begin{align} pp
&\rightarrow p \,(\gamma\gamma
\rightarrow \gamma\gamma) \,p \;, \label{pp_collision} \\
p Pb &\rightarrow p \,(\gamma\gamma
\rightarrow \gamma\gamma) \,Pb \;, \label{pN_collision}  \\
Pb Pb &\rightarrow Pb \,(\gamma\gamma \rightarrow \gamma\gamma) \,Pb
\;. \label{NN_collision}
\end{align}
The FCC-hh, operating at $100$ TeV center-of-mass energy, provides a
unique opportunity to significantly extend the search for heavy
ALPs. In addition to proton-proton collisions, the availability of
proton-ion and ion-ion collision modes allows one to exploit
enhanced photon fluxes due to the large nuclear charge, effectively
turning these systems into photon-photon colliders.

This paper presents a systematic and comparative
analysis of ALP contributions to LbL scattering in pp, pPb, and PbPb
collisions at the FCC-hh. By examining these collision modes within
a unified framework, we aim to highlight their complementary
sensitivities across different ALP parameter space regions.

The main characteristics of $\gamma\gamma\rightarrow\gamma\gamma$
processes at the FCC-hh are given in Tab.~\ref{tab:1}. For
comparison, pp, pPb and PbPb collisions at the LHC are studied to
date at a center-of-mass energy per nucleon of 13 TeV, 8.16 TeV, and
5.02 TeV, respectively.
%
\begin{table}
\begin{center}
\begin{tabular}{|l|c|c|c|c|c|}
  \hline
  Collision & $\sqrt{s_{N\!N}}$ & $\mathcal{L}$ & $\gamma_L$ & $\omega_{\max}$ & $\sqrt{s^{\max}_{\gamma\gamma}}$ \\
      & (TeV) & (per year) & ($\times 10^3$) & (TeV) & (TeV) \\
  \hline
  pp & 100 & 1 fb$^{-1}$ & 53.0 & 17.6 & 35.2 \\
  pPb & 62.8 & 8 pb$^{-1}$ & 33.5 & 0.95 & 1.9 \\
  PbPb & 39.4 & 33 nb$^{-1}$ & 21.0 & 0.60 & 1.2 \\
  \hline
\end{tabular}
\end{center}
\caption{Characteristics of $\gamma\gamma\rightarrow\gamma\gamma$
processes at the FCC-hh. Here $\sqrt{s_{N\!N}}$ is the
nucleon-nucleon c.m. energy, $\mathcal{L}$ is the integrated
luminosity per year (a ``year'' is $10^7$ s for pp collisions, and
$10^6$ s for the ion mode), $\gamma_L$ is the Lorentz factor,
$\omega_{\max}$ is the maximum photon energy in the c.m.s., and
$\sqrt{s^{\max}_{\gamma\gamma}}$ is the maximum invariant energy of
the photon-photon system \protect{\cite{FCC:physics}}.}
\label{tab:1}
\end{table}

The coherent contribution of $Z$ protons in a heavy-ion collision provides
an intense spectrum of equivalent photons in the ions with nuclear
charge $Z$. Since each photon flux scales as the square of the ion
charge $Z^2$, the $\gamma\gamma$ luminosity in pPb (PbPb) collision
is extremely enhanced by a factor of $Z^2$ ($Z^4$) compared to
similar pp interactions. This compensates the lower integrated
luminosity compared with that available in pp collision. Thus, both
the LHC and FCC-hh operating in ion modes can be considered as
photon-photon colliders. The LbL scattering in ion collisions at the
LHC was experimentally studied in \cite{ATLAS_LbL_axion:2021},
\cite{CMS_LbL_axion:2025}, \cite{ATLAS_ions_1}-\cite{CMS_ions_2}.
Phenomenology of the LbL scattering at the LHC and future colliders
was addressed in \cite{I-K:2020,I-K:2021}, \cite{I-K:2023},
\cite{Beldenegro:2018}, \cite{Coelho:2020_2},
\cite{Atag:2010,Inan:2019}. The FCC-hh, due to the higher diphoton
pairs reached, may be sensitive to very heavy ALPs contributing to
the LbL scattering.

The novelty of this study lies in the systematic comparison of the
three collision modes within a unified framework, highlighting their
complementary sensitivities across different mass regions. We show
that heavy-ion collisions provide enhanced sensitivity in the
intermediate mass range due to the strong photon flux, while
proton-proton collisions dominate at higher masses thanks to their
larger kinematic reach. This interplay allows the FCC-hh to probe
ALP parameter space well beyond current LHC limits, particularly in
the TeV-scale mass region.

\section{LbL virtual production of ALP} %

The Lagrangian of the pseudoscalar ALP (in what follows, denoted as
$a$) looks like
\begin{equation}\label{axion_photon_lagrangian}
\mathcal{L}_a = \frac{1}{2}\,(\partial_\mu a)(\partial^\mu a) -
\frac{1}{2} m_a^2 a^2 - \frac{1}{f_a} \,aF_{\mu\nu}
\tilde{F}^{\mu\nu} \;,
\end{equation}
where $m_a$ is the ALP mass, $F_{\mu\nu}$ is the energy-momentum
tensor of the photon field, $\tilde{F}_{\mu\nu} = (1/2)
\varepsilon_{\mu\nu\rho\sigma} F^{\rho\sigma}$ its dual, and $f_a$
is the ALP-photon coupling.

Then the total width of the ALP is equal to
\begin{equation}\label{axion_width}
\Gamma_a =
\frac{\Gamma(a\rightarrow\gamma\gamma)}{\mathrm{Br}(a\rightarrow\gamma\gamma)}
\;,
\end{equation}
where
\begin{equation}\label{photon_axion_width}
\Gamma(a\rightarrow\gamma\gamma) = \frac{m_a^3}{4\pi f_a^2}
\end{equation}
is its total decay width and
${\mathrm{Br}(a\rightarrow\gamma\gamma)}$ is the branching ratio for
decaying into two photons.

In the equivalent photon approximation (EPA)
\cite{Budnev:1975,Baur:2002} differential cross sections of the
processes \eqref{pp_collision}-\eqref{NN_collision} can be
factorized as
\begin{equation}\label{diff_cs}
d\sigma_{AB} = \int\limits_{W_{\min}}^{W_{\max}}
\frac{dL_{\gamma\gamma}}{dW} \, d\hat{\sigma}_{\gamma\gamma \to
\gamma\gamma}(W)\, dW \;,
\end{equation}
where  $A,B = p \ \mathrm{or}\, N$, $W$ is the center of mass energy
of the two photon system, and
\begin{equation}\label{L}
\frac{dL_{\gamma\gamma}}{dW} = \frac{W}{2}
\int\limits_{\omega_{\min}}^{\omega_{\max}}
\!\!f_{\gamma/A}\!\left(\frac{W^2}{4\omega}\right)
\!f_{\gamma/B}(\omega) \,\frac{d\omega}{\omega}
\end{equation}
is the $\gamma\gamma$-luminosity. Here $f_{\gamma/A}(\omega) =
dN_A/d\omega$ is the photon distributions in energy $\omega$ inside
the proton or nucleus. Let $x_1$ and $x_2$ be energy fractions of
the colliding photons, $x_i = \omega_i/E_B$, where $E_B$ is the beam
energy. The invariant energy is defined as $W = 2E_p\sqrt{x_1 x_2}$
and $W = 2E_N\sqrt{x_1 x_2}$ for the symmetric proton-proton and
lead-lead collisions, respectively, and $W = 2\sqrt{E_p
E_N}\sqrt{x_1 x_2}$ for the pPb scattering. The integration limits
in eq.~\eqref{L} are given by
\begin{equation}\label{y_limits}
\omega_{\min} = \max\!\left(\frac{W^2}{4\xi_{\max}E_B}, \; x_{\min}
E_B \right), \qquad \omega_{\max} = x_{\max}E_B \;,
\end{equation}
where
\begin{equation}\label{ksi_limits}
x_{\max} = 1 - \frac{m_{p/N}}{E_B} \;, \qquad x_{\min} =
\frac{p_T}{E_B} \;,
\end{equation}
$m_{p/N}$ is the mass of a colliding particles, and $p_T$ is the
transverse momenta of the outgoing photons. Correspondingly, the
lower and upper limits on the invariant energy of the photon pair in
\eqref{diff_cs} are equal to
\begin{equation}\label{W_limits}
W_{\min} = 2E_B x_{\min} \;, \qquad W_{\max} = 2E_B x_{\max} \;.
\end{equation}

\subsection{Photon distribution in proton} %

In the EPA \cite{Budnev:1975} the photon distribution for photons
which are emitted from a proton beam is given through the formula
\cite{Kepka:2008,Sahin:2013}
\begin{equation}
f_{\gamma/p}(\omega) = \frac{\alpha}{\pi \omega} \left(1 -
\frac{\omega}{E_B}\right) \!\left[
\phi\!\left(\frac{Q_{\max}^2}{Q_0^2}\right) -
\phi\!\left(\frac{Q_{\min}^2}{Q_0^2}\right) \right] ,
\end{equation}
where the function $\phi(x)$ is defined as
\begin{align}\label{phi}
\phi(x) &= (1 + a y) \left[ -\ln\!\left(1 + \frac{1}{x}\right) +
\sum_{k=1}^{3} \frac{1}{k(1+x)^{k}} \right] +
\frac{y(1-b)}{4x(1+x)^{3}}
\\
&\quad + c\left(1 + \frac{y}{4}\right) \left[
\ln\!\left(\frac{1-b+x}{1+x}\right) + \sum_{k=1}^{3}
\frac{b^{k}}{k(1+x)^{k}} \right] ,
\end{align}
with $Q_{\min}^2 = 0.71$ GeV$^2$ and $Q_{\max}^2 = 2$ GeV$^2$. Here
\begin{equation}\label{y}
y = \frac{\omega^2}{E_B(E_B - \omega)} \;,
\end{equation}
with the parameters $a,b,c$ given by
\begin{align}
a &= 1 + \frac{\mu_p^{2}}{4} + \frac{4m_p^{2}}{Q_0^{2}} \approx 7.16
\;,
\\ \nonumber
b &= 1 - \frac{4m_p^{2}}{Q_0^{2}} \approx -3.96 \;,
\\ \nonumber
c &= \frac{\mu_p^{2} - 1}{b^{4}} \approx 0.028 \;,
\end{align}
where $\mu_p \approx 2.793$ is the anomalous magnetic moment of the
proton, and $m_p$ its mass.

\subsection{Photon distribution in nucleus} %

We consider an exclusive production of two photons in a heavy-ion
collision with the diphoton final-state measured in the central
detector, and nuclei surviving the electromagnetic interaction
scattered at very low angles with respect to the beams. In the EPA
\cite{Budnev:1975} the accelerated ions can be considered as
$\gamma$ beams. Indeed, the emitted photons are almost on-shell,
since their virtuality $|Q^2| < 1/R_A^2$, where $R_A = 1.2A^{1/3}$
fm is the radius of the nucleus. In the case of a lead-lead
collision with A = 208, it results in $|Q^2| < 7.7 \cdot 10^{-4}$
GeV$^2$.

In the relativistic limit the equivalent spectrum of the photon from
the nucleus $N$ with the charge $Z$ and atomic number $A$ is given
by \cite{Baur:2002}, \cite{Jackson:QED}-\cite{Baltz:2008}
\begin{equation}\label{dist_gamma_N}
f_{\gamma/N}(\omega) = \frac{2Z^2\alpha}{\pi\omega} \!\left[ \xi
K_0(\xi) K_1(\xi) - \frac{\xi^2}{2} ( K_1^2(\xi) - K_0^2(\xi) )
\right] ,
\end{equation}
where $\xi = \omega/E_R$, $E_R = E_N/(m_N R_A) = \sqrt{s_{NN}}/(2m_p
R_A)$. $K_0(x)$ ($K_1(x)$) is the modified Bessel function of the
second kind of order zero (one).

\subsection{LbL helicity amplitudes} %

The differential cross section of the process
$\gamma\gamma\rightarrow\gamma\gamma$ (after accounting for the
$P$-parity and Bose-Einstein symmetry) is given by the following sum
of helicity amplitudes squared \cite{Beldenegro:2018}
\begin{equation}\label{diff_cs_ampl}
\frac{d\sigma}{d\Omega} = \frac{1}{128\pi^2 s} \left( |M_{++++}|^2 +
|M_{+-+-}|^2 + |M_{+--+}|^2 + |M_{++--}|^2 \right) .
\end{equation}
Here and below $s$, $t$ and $u$ are the Mandelstam variables of the
diphoton system. Each of the helicity amplitudes is a sum of ALP and
SM terms,
\begin{equation}\label{ALP+SM}
M = M^{(a)} + M_{\mathrm{SM}} \;.
\end{equation}

The explicit expressions of the pure ALP amplitudes can be found in
\cite{Beldenegro:2018}:
\begin{align}\label{M_hel_ampl}
M_{++++}^{(a)} &= - \frac{4}{f_a^2} \frac{s^2}{s - m_a^2
+ im_a\Gamma_a} \;, \nonumber \\
M_{+-+-}^{(a)} &= - \frac{4}{f_a^2} \frac{u^2}{u - m_a^2 + im_a\Gamma_a} \;, \nonumber \\
M_{+--+}^{(a)} &= - \frac{4}{f_a^2} \frac{t^2}{t - m_a^2 +
im_a\Gamma_a} \;, \nonumber \\
M_{++--}^{(a)} &=  \frac{4}{f_a^2} \left( \frac{s^2}{s - m_a^2 +
im_a \Gamma_a} + \frac{t^2}{t - m_a^2 + im_a\Gamma_a} \right. \nonumber \\
&+ \left. \frac{u^2}{u - m_a^2 + im_a\Gamma_a} \right) .
\end{align}

In the LO approximation each of the SM amplitudes of the
$\gamma\gamma\rightarrow\gamma\gamma$ process is a sum of fermion
and $W$ boson one-loop amplitudes, $M_{\mathrm{SM}} = M_f + M_W$.
The analytic (rather complicated) expressions for one-loop
amplitudes $M_f$ an $M_W$ have been obtained in
\cite{Jikia:1994}-\cite{Gounaris:1999_2} (see also\cite{Atag:2010}).
As shown in \cite{Bern:2001}-\cite{Ajjath:2024_2}, NLO (QCD + QED)
corrections to the LbL cross section are quite small numerically (a
few percent at $p^\gamma_T > 5$ GeV and $|\eta^\gamma| \leq 2.4$),
showing that the LO computations are robust.

In Fig.~\ref{fig:FCCallE50m} cross sections of the photon pair
production in PbPb, pPb and pp collisions at the FCC-hh are shown.
The total cross sections (upper curves in these figures) account for
contributions from ALP through the
$\gamma\gamma\rightarrow a \rightarrow \gamma\gamma$ scattering.

\section{Numerical results} %

\begin{figure}[htb]
\begin{center}
\includegraphics[scale=0.5]{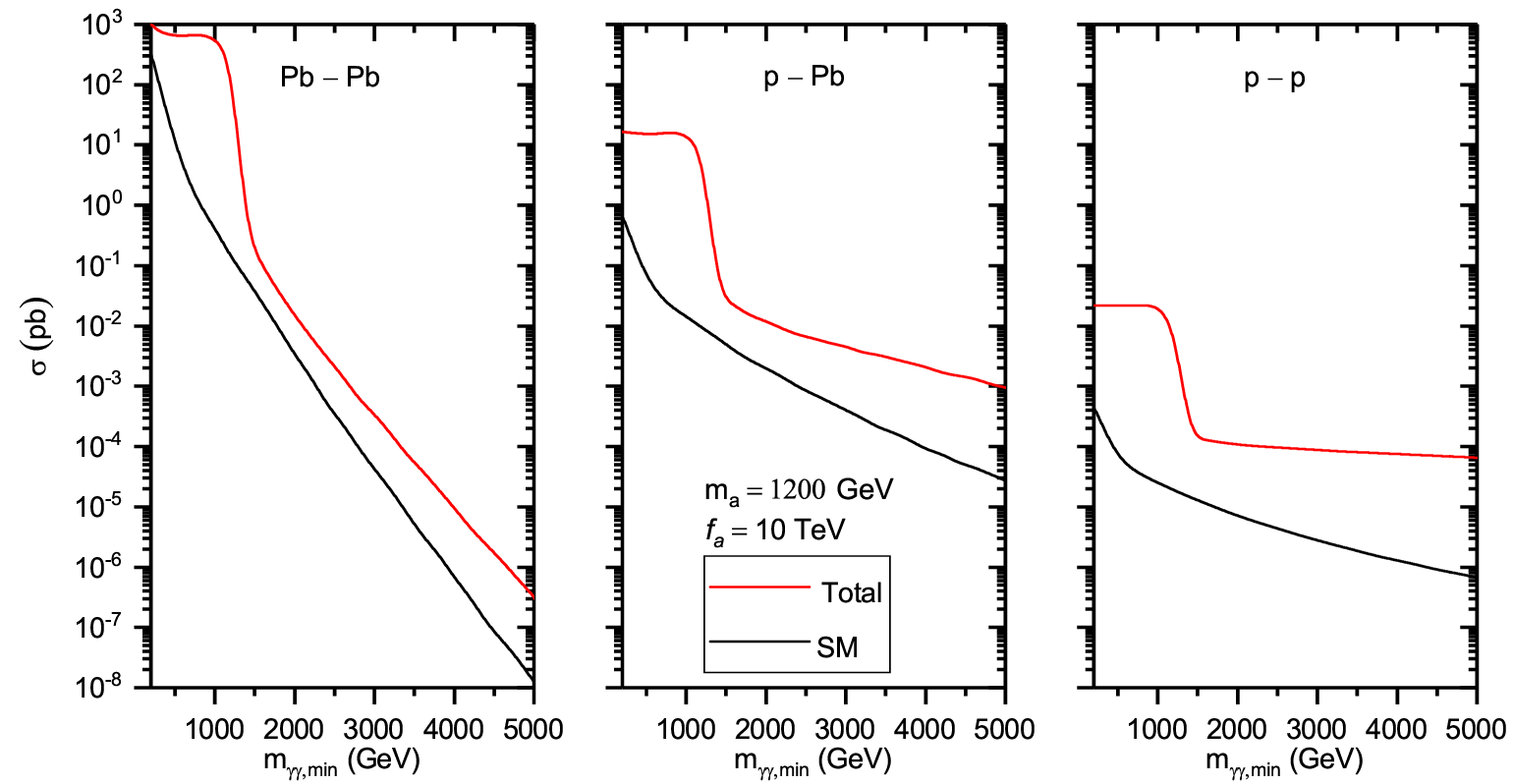}
\caption{The total and SM cross sections for the processes
$Pb(\gamma\gamma\rightarrow\gamma\gamma)Pb$ (left panel),
$p(\gamma\gamma \rightarrow\gamma\gamma)Pb$ (middle panel), and
$p(\gamma\gamma\rightarrow\gamma\gamma)p$ (right panel) at the
FCC-hh depending on the invariant mass of the outgoing photons. The
ALP mass is $m_a = 1.2$ TeV, and the ALP-photon coupling is
$f_a^{-1} = 0.1$ TeV$^{-1}$.}
\label{fig:FCCallE50m}
\end{center}
\end{figure}

Using ALP and SM helicity amplitudes, we have calculated
differential cross sections for the pp, pPb and PbPb collisions,
accounting for contribution from ALPs. They are presented by upper
curves in Fig.~\ref{fig:FCCallE50m} depending on the minimal value
of the invariant mass of the outgoing photons $m_{\gamma\gamma}$.
The lower curves in Fig.~\ref{fig:FCCallE50m} correspond to the SM
cross sections. As expected, the PbPb cross section is the largest
one up to $m_{\gamma\gamma} \approx 1.2$ TeV, but it falls most
rapidly as $m_{\gamma\gamma}$ grows.

\begin{figure}[htb]
\begin{center}
\includegraphics[scale=0.55]{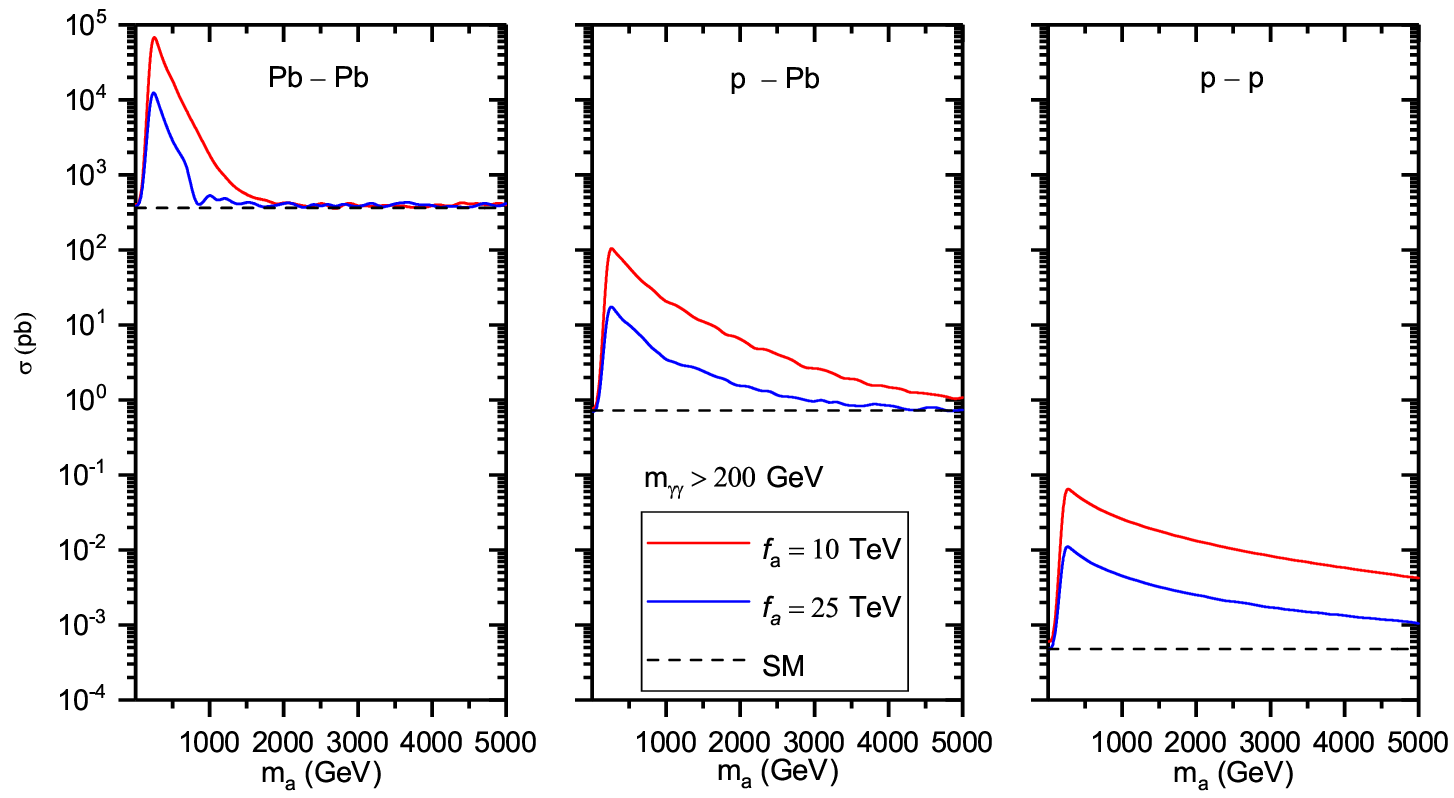}
\caption{The total cross sections for the processes
$Pb(\gamma\gamma\rightarrow\gamma\gamma)Pb$ (left panel),
$p(\gamma\gamma \rightarrow\gamma\gamma)Pb$ (middle panel), and
$p(\gamma\gamma\rightarrow\gamma\gamma)p$ (right panel) at the
FCC-hh depending on the ALP mass. The ALP-photon coupling constant
is taken to be $f_a^{-1} = $ TeV$^{-1}$. The cut $m_{\gamma\gamma} >
200$ GeV is imposed.}
\label{fig:FCCallE50fm_a}
\end{center}
\end{figure}

In Fig.~\ref{fig:FCCallE50fm_a} the total cross sections are given
depending on the ALP mass $m_a$ in the range 1 GeV -- 5000 GeV. The
figure shows that an ALP contribution to the cross sections tends to
zero with an increase of $m_a$. This effect is particularly
pronounced for the ion-ion collisions and it is postponed for the
proton-proton collisions. This is due to the explicit dependence of the
helicity amplitudes on Mandelstam variables \eqref{M_hel_ampl} and
rapid growth of the ALP total width with $m_a$
\eqref{photon_axion_width}. The different size of the effect is
explained by different values of the invariant energy of the
photon pair $\sqrt{s^{\max}_{\gamma\gamma}}$ in PbPb, pPb and pp
collisions, see the last column in Tab.~\ref{tab:1}.

Let $S$ and $B$ be the total number of signal and SM events, and
$\delta$ is the percentage systematic uncertainty. The exclusion
significance $SS_{\mathrm{excl}}$ is known to be
\cite{Cowan:2011}-\cite{Zhang:2020}
\begin{align}\label{S_excl}
SS_{\mathrm{excl}} = \bigg\{ &2 \left[ S - B \ln\!\left( \frac{B + S
+ x}{2B} \right) - \frac{1}{\delta^2}\ln\!\left( \frac{B - S +
x}{2B}
\right) \right] \nonumber \\
&- (B + S - x) \left( 1 + \frac{1}{\delta^2B} \right) \bigg\}^{1/2}
,
\end{align}
with
\begin{equation}\label{x}
x = \sqrt{ (S + B)^2 - \frac{4S\delta^2B^2}{(1 + \delta^2 B)} } \;.
\end{equation}
Correspondingly, the discovery significance $SS_{\mathrm{disc}}$ is
defined as \cite{Cowan:2011}-\cite{Zhang:2020}
\begin{align}\label{S_disc}
SS_{\mathrm{disc}} = \bigg\{ &2\left[ (S + B) \ln\!\left( \frac{(B +
S)(1 + \delta^2 B)}{B + \delta^2 B(S + B)} \right) \right] \nonumber
\\
&- \frac{1}{\delta^2}\ln\!\left( 1 + \frac{\delta^2 S}{1 + \delta^2
B} \right) \bigg\}^{1/2} \;.
\end{align}
We classify the region $SS_{\mathrm{excl}} \leqslant 1.645$ as an
exclusion region at the 95\% confidence level, whereas
$SS_{\mathrm{disc}}
> 3(5)$ as a discoverable region at 3\,$\sigma$(5\,$\sigma$).

\begin{figure}[htb]
\begin{center}
\includegraphics[scale=0.55]{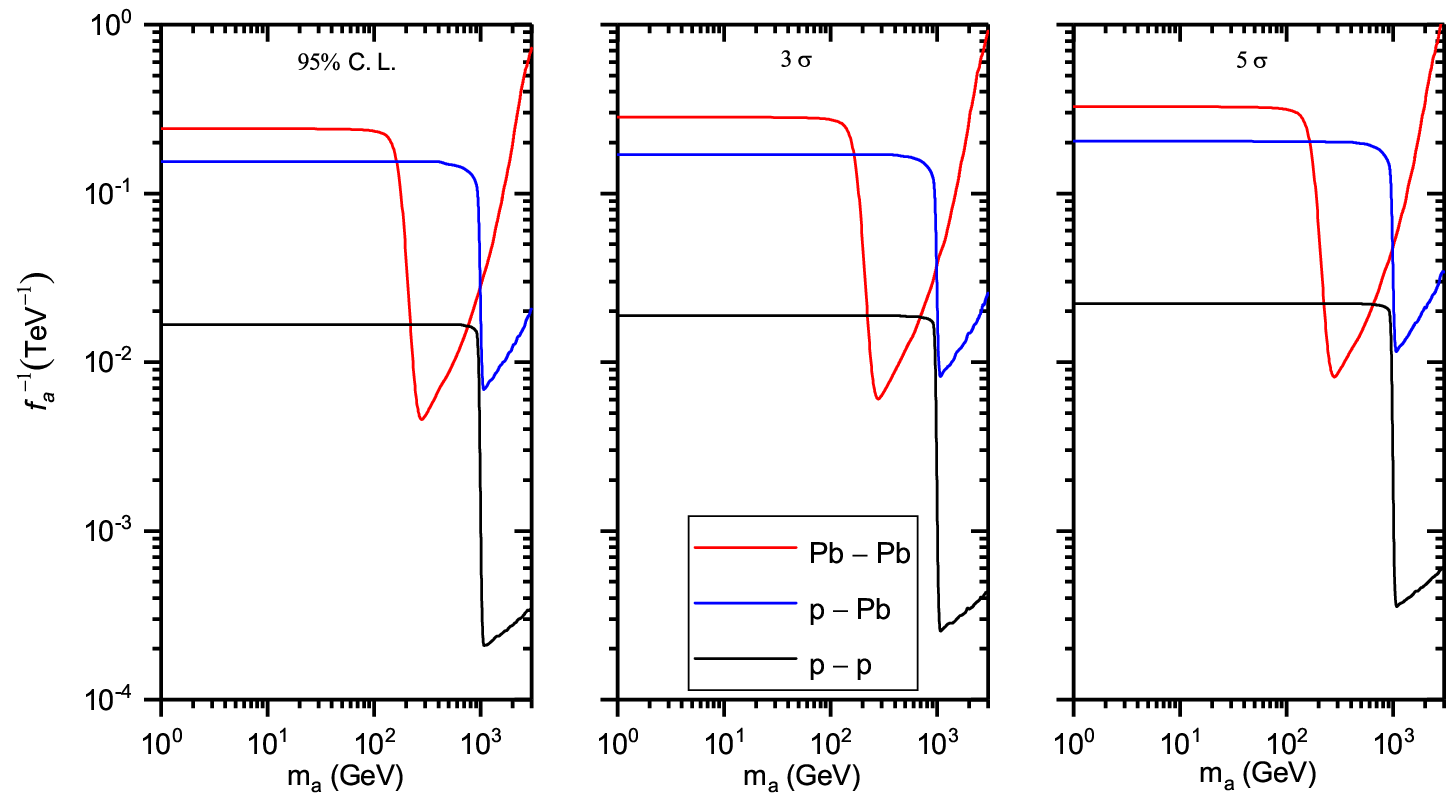}
\caption{The 95\% C.L. limits and $3\sigma (5\sigma)$ discovery
limits on the ALP coupling constant depending on the ALP mass from
$Pb(\gamma\gamma\rightarrow\gamma\gamma)Pb$, $p(\gamma\gamma
\rightarrow\gamma\gamma)Pb$, and
$p(\gamma\gamma\rightarrow\gamma\gamma)p$ collisions.}
\label{fig:FCCSSall_delta_0}
\end{center}
\end{figure}

The 95\% C.L. limits on the ALP coupling are shown in the left panel
of Fig.~\ref{fig:FCCSSall_delta_0} versus the ALP mass in the mass
range 1 GeV -- 3 TeV for PbPb, pPb and pp collisions. For the
integrated luminosity in pp collisions we have used
$\mathcal{L}_{pp} = 30$ ab$^{-1}$. For pPb and PbPb scattering, we
have taken the values of $\mathcal{L}_{pPb} = 27$ pb$^{-1}$ and
$\mathcal{L}_{PbPb} = 110$ nb$^{-1}$, respectively, following
refs.~\cite{FCC:V_3} and \cite{Dainese:2017, Abraham:2025}.

The systematic uncertainty $\delta$ in eqs.~(21), (23) effectively
parametrizes uncertainties from a luminosity, photon flux modeling,
detector efficiency, and background estimations. We adopt a
representative value of $\delta = 5-10\%$, as commonly done in
similar collider studies.

In our case, the background is strongly suppressed ($B\ll S$). In a
small-background limit the exclusion significance
$SS_{\mathrm{excl}}$ is estimated as
\begin{equation}\label{S_excl_exp}
SS_{\mathrm{excl}} = \sqrt{2S} \left( 1 + \frac{\varepsilon \ln
\varepsilon}{2} \right) + \mathrm{O}(\varepsilon^2) \;,
\end{equation}
where $\varepsilon = B/S$. Thus, the dependence on the uncertainty
completely disappears in this limit. As for the discovery
significance $SS_{\mathrm{disc}}$, if $B\ll S, B\delta^2 < 1$ the
dependence on $\delta$ enters only through the combination
$B\delta^2$, and appears inside a logarithm,
\begin{equation}\label{S_excl_exp}
SS_{\mathrm{disc}} = \sqrt{ 2S \ln\left(\frac{1}{\varepsilon +
B\delta^2}\right) } + \mathrm{O}(\varepsilon) \;.
\end{equation}
This implies that the dependence on the systematic uncertainty is
still present in $SS_{\mathrm{disc}}$, but it is relatively weak.
Our numerical calculations confirm this statement.

The behavior of the exclusion regions in the left panel of
Fig.~\ref{fig:FCCSSall_delta_0}, namely, specifically different
plateau lengths and positions of the dips are directly coupled to
how each collision mode interacts with the $m_{\gamma\gamma} > 1000$
GeV threshold.

The plateau represents a "massless-limit" regime where the
$m_{\gamma\gamma}$ cut is much larger than the ALP mass $m_a$:
\begin{description}
  \item[--]
For pp and pPb collisions the plateaus extend up to nearly 1000
GeV and 900 GeV, respectively.

This is because the protons give a very ``hard'' photon spectrum.
Even as $m_a$ begins to approach the 1 TeV, the signal yield above the
1000 GeV threshold remains dominated by the high-energy
photon-photon luminosity, keeping the limit on $f_a^{-1}$ constant.
  \item[--]
For the PbPb case, the plateau is much shorter, ending around 180
GeV.

This earlier transition occurs because a heavy-ion photon flux is
considerably ``softer'' than that in the proton cases, and the
sensitivity starts to change much earlier as $m_a$ grows. The energy
required to produce the ALP and simultaneously satisfy the 1 TeV cut
quickly reaches the limit of what the lead nucleus can provide
coherently.
\end{description}

The origin and positions of the ``dips'' in the left panel of
Fig.~\ref{fig:FCCSSall_delta_0} can be explained by the following:
\begin{description}
  \item[--]
For the pp and pPb collisions the dips of the figure occur around
1000 GeV. This is a ``resonance'' effect where the ALP mass matches
the di-photon mass threshold. At this point, the cross-section is
most efficiently sampled right to the $m_{\gamma\gamma}$ cut
imposed.
  \item[--]
For the PbPb mode the dip occurs much earlier.

Due to the nuclear form factor, the photon flux is heavily
suppressed at high energy scales. We find a ``sweet spot'' around
200 GeV, where the $Z^4$-enhancement remains highly effective, while
the energy is still sufficient to satisfy the $1$ TeV invariant mass
threshold. Beyond this point, the form factor suppression outweighs
the $Z^4$-enhancement, causing the exclusion limit to weaken.
\end{description}

The fact that the PbPb collision is more sensitive to the ALP
coupling in the $60-700$ GeV mass range shows that the photon flux
enhancement is powerful enough to overcome the luminosity deficit
($110 \text{ nb}^{-1}$ against $20 \text{ ab}^{-1}$), but only
within a range where the ion-ion scattering remains coherent.

The $3\sigma$ and $5\sigma$ discovery limits on the ALP coupling are
presented in the middle and right panels of
Fig.~\ref{fig:FCCSSall_delta_0}, respectively. As is the case with
the exclusion constraints, the most stringent $3\sigma$ ($5\sigma$)
limits on the ALP coupling $f_a^{-1}$ are achieved for the
proton-proton collision around $m_a \simeq 1$ TeV.

A probe of the ALP production in pp and pPb collisions at the FCC-hh
could be realized by searching for $\gamma\gamma$ pairs produced in
coincidence with intact protons reconstructed in very forward proton
detector (FPD), such as the Precision Proton Spectrometer planned by
the CMS Collaboration for the HL-LHC
\cite{CMS_PPS_1}-\cite{CMS_PPS_3}. In a central exclusive production
like $pp \rightarrow pXp$ or $pPb \rightarrow pXPb$ intact protons
lose a fraction $\xi = \Delta p/p$ of their longitudinal momentum.
We assume that the FPD at the FCC-hh could have an acceptance of
$0.003 \leq \xi \leq 0.2$.

\begin{figure}[htb]
\begin{center}
\includegraphics[scale=0.55]{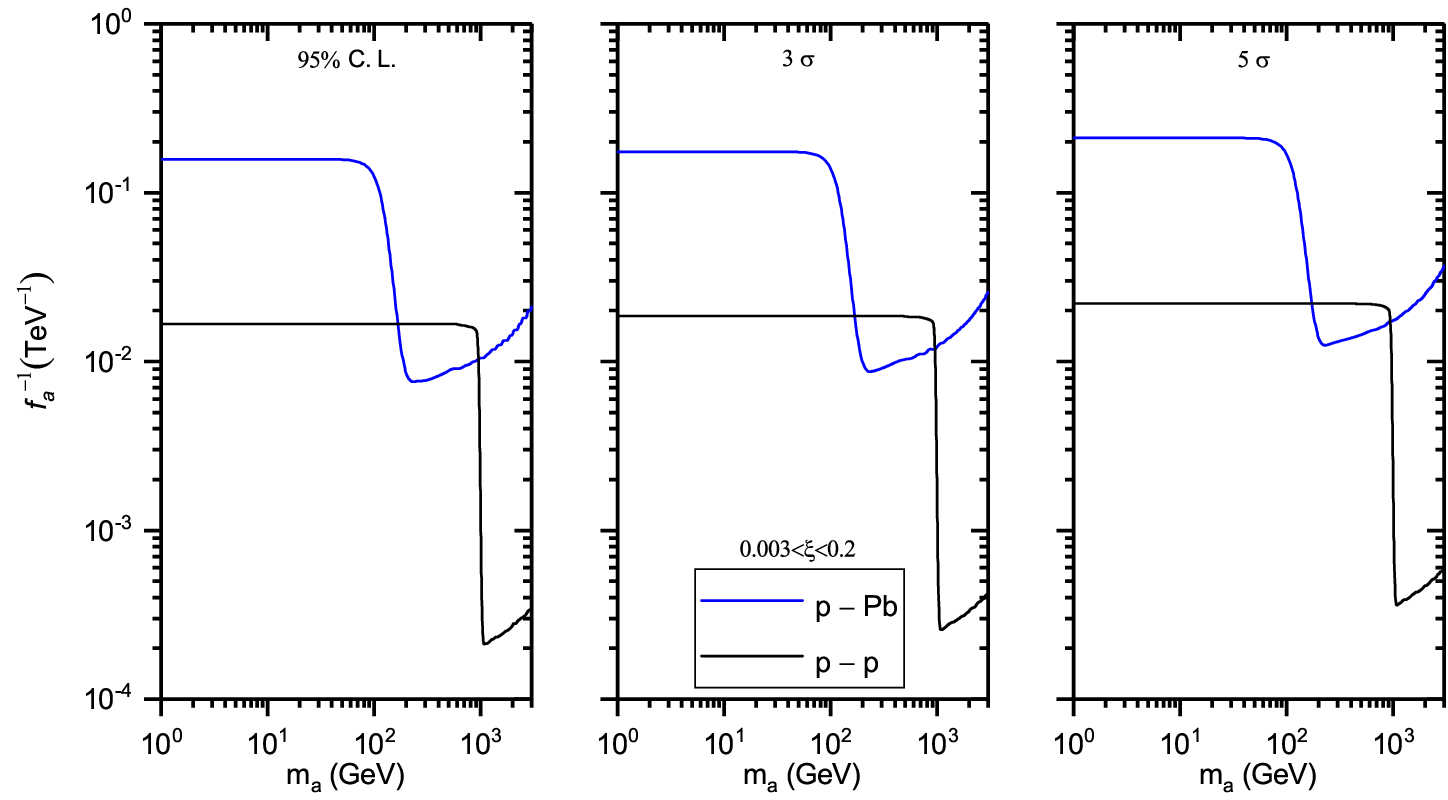}
\caption{The 95\% C.L. limits and $3\sigma (5\sigma)$ discovery
limits on the ALP coupling constant depending on the ALP mass for
the $p(\gamma\gamma \rightarrow\gamma\gamma)Pb$ and
$p(\gamma\gamma\rightarrow\gamma\gamma)p$ collisions, with the
proton tagging at the FCC-hh.} \label{fig:FCCSSall_02_003_delta_0}
\end{center}
\end{figure}

The 95\% C.L. limits, as well as $3\sigma$ and $5\sigma$ discovery
limits on the ALP coupling obtained with the use of the FPD are
shown in Fig.~\ref{fig:FCCSSall_02_003_delta_0}. To improve the
ratio $S/B$, we have imposed the cut $m_{\gamma\gamma} > 1$ TeV. The
invariant mass of the outgoing diphoton pair is equal to
$\sqrt{\xi_1\xi_2 s}$, where $\xi_1$ and $\xi_2$ are the energy
fractions of the colliding photons. At the 100 TeV hadron collider,
the lower bound $\xi_{\min} = 0.003$ sets a threshold of
$m_{\gamma\gamma} \simeq 236$ GeV. To obey the cut $m_{\gamma\gamma}
> 1$ TeV, it is enough to have a FPD acceptance of $\xi > 0.01$.
Thus, the value of the pp cross section does not change, if we use
$\xi_{\min} = 0.003$. Moreover, the dominant contribution to the
cross section comes from an intermediate region $\xi \simeq
0.01-0.1$, i.e. well below 0.2. Therefore, the upper bound
$\xi_{\max} = 0.2$  does not effectively remove any relevant part of
the phase space, and the bounds on $f_a^{-1}$ remain essentially
unchanged.

In contrast, in the pPb collision a situation is quite different
because of a much softer spectrum of the photon emitted by the lead
ion \eqref{dist_gamma_N}, which makes the cross sections sensitive
to the $\xi$ cuts. As a result, we come to stronger bounds on
$f_a^{-1}$ in the ALP mass region 80 GeV -- 1 TeV. This behavior can
be verified by comparing the curves in
Fig.~\ref{fig:FCCSSall_02_003_delta_0} with the corresponding curves
in Fig.~\ref{fig:FCCSSall_delta_0}.

\section{Conclusions} %

In the present paper we have examined the contribution of massive
axion-like particles to light-by-light scattering in proton-proton,
proton-lead and lead-lead collisions at the collider FCC-hh. Our
main results are the 95\% C.L. exclusion limits and
$3\sigma$($5\sigma$) discovery limits on the ALP coupling constant
$f_a^{-1}$ as a function of the ALP mass $m_a$. We have used the
integrated luminosities of 30 ab$^{-1}$, 27 pb$^{-1}$ and 110
nb$^{-1}$ for pp, pPb and Pb collisions, respectively. The strongest
limit on the ALP coupling is obtained in the mass region close to
$m_a \simeq 250$ GeV for the PbPb collisions and $m_a \simeq 1$ TeV
for pp and pPb processes.

A primary result of this study is the clear evidence of the
complementarity between these collision modes at the FCC-hh. While
PbPb collisions benefit from a strong enhancement of the photon flux
due to the large nuclear charge, making them particularly sensitive
in the intermediate mass region, pp case provides access to higher
invariant masses and thus dominates the sensitivity in the TeV-scale
region. Hence, the results show that the FCC-hh can significantly
extend current bounds on the ALP-photon coupling obtained at the
LHC, especially in the high-mass region.

A search for ALPs has been carried out at the LHC with the ATLAS
experiment using 14.6 fb$^{-1}$ of $\sqrt{s} = 13$ TeV pp collisions
\cite{ATLAS_pp_axion:2023}. Events with centrally produced photon
pairs tagged by forward scattered protons in $pp \rightarrow
p(\gamma\gamma \rightarrow \gamma\gamma)p^{(*)}$ scattering,
mediated by an ALP resonance, have been studied. The search was
performed in the diphoton mass range $m_{\gamma\gamma}$ from 150 to
1600 GeV. The upper limit on the ALP coupling constant, assuming a
100\% decay branching ratio into two photons,
\begin{equation}\label{ATLAS_bound_pp}
f_a^{-1} \leq 0.04 - 0.09  \  \mathrm{TeV}^{-1} \;,
\end{equation}
was obtained at the 95\% C.L. \cite{ATLAS_pp_axion:2023}.

The CMS and TOTEM Collaborations have also performed a search
for events with a high-mass exclusive diphoton system and two intact
protons in the final state in $pp$ collisions at $\sqrt{s} = 13$
TeV, with the integrated luminosity of 103 fb$^{-1}$
\cite{CMS_pp_axion:2024}. Events that have two photons with high
transverse momenta ($p_T > 100$ GeV), back-to-back in azimuth, and
large diphoton invariant mass ($m_{\gamma\gamma} > 350$ GeV)
selected. As a result, the limits on the production of ALPs coupling
to photons with strengths
\begin{equation}\label{CMS_bound_pp}
f_a^{-1} \approx  0.03 \ \mathrm{TeV}^{-1}
\end{equation}
were set over the ALP mass range from 500 to 2000 GeV
\cite{CMS_pp_axion:2024}. More accurately, $f_a^{-1}$ varies from
0.03 TeV$^{-1}$ to 1 TeV$^{-1}$ over this mass range
\cite{CMS_pp_axion:2024}. This limit is the most restrictive LHC
bound to date on the ALP-photon coupling, in the very high mass
region.

For the low mass range ($5 - 100$ GeV), the constraints on the coupling
$f_a^{-1}$ have been obtained in the PbPb collisions at the LHC. The
ATLAS collaboration examined the LbL scattering events with two
photons produced exclusively, each with a transverse energy
$E_T^\gamma > 2.5$ GeV, a pseudorapidity $|\eta^\gamma| < 2.37$, a
diphoton invariant mass $m_{\gamma\gamma} > 5$ GeV, and with a small
diphoton transverse momentum and diphoton acoplanarity
\cite{ATLAS_LbL_axion:2021}. The measurements were based on the data
with an integrated luminosity of 2.2 nb$^{-1}$. The limits
\begin{equation}\label{ATLAS_bound_ions}
f_a^{-1} = 0.06 - 0.3 \ \mathrm{TeV}^{-1}
\end{equation}
have been set in the ALP mass range $6-100$ GeV.

\begin{figure}[htb]
\begin{center}
\hspace*{-0.4cm}
\includegraphics[scale=0.35]{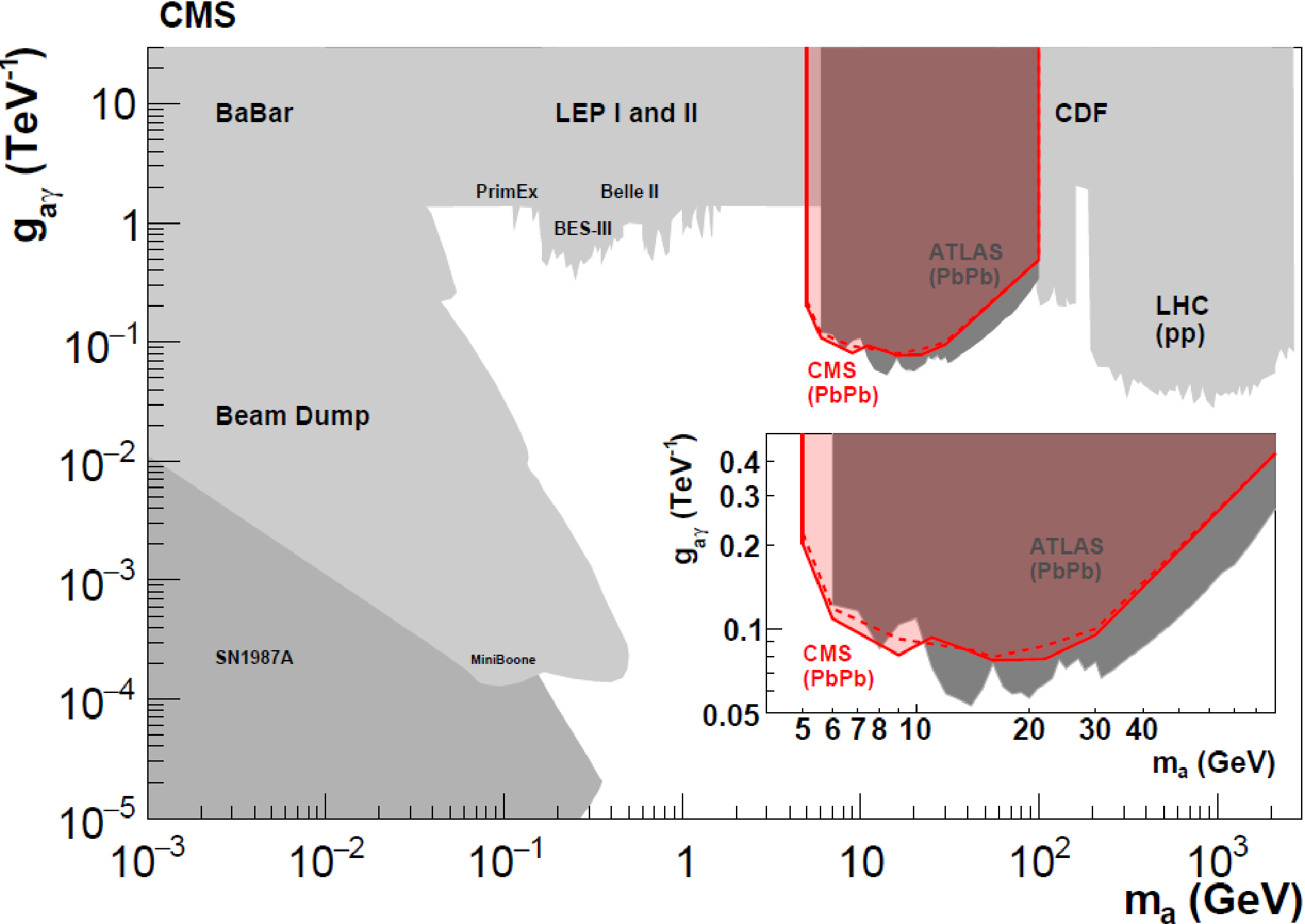}
\caption{The exclusion limits at 95\% CL on the ALP-photon coupling
$g_{a\gamma}$ versus ALP mass $m_a$ plane, for the operator $(1/4)
aF\tilde{F}$ (assuming ALPs coupled only to photons)
\protect{\cite{CMS_LbL_axion:2025}}.}
\label{fig:ALP_coupling_vs_mass}
\end{center}
\end{figure}

Measurements of the LbL scattering in ultraperipheral collisions of
lead ions at the LHC with an integrated luminosity of 1.7 nb$^{-1}$
were reported by the CMS Collaboration. The collisions were required
to have a transverse energy of $E^\gamma_T > 2$ GeV, a
pseudorapidity of $|\eta^\gamma| < 2.2$, and the pairs to have an
invariant mass of $m_{\gamma\gamma} > 5$ GeV, a transverse momentum
of $p^{\gamma\gamma}_T < 1$ GeV, and an azimuthal $\gamma\gamma$
acoplanarity of 0.1 \cite{CMS_LbL_axion:2025}. The couplings larger
than
\begin{equation}\label{CMS_bound_ions}
f_a^{-1} = 0.1 - 0.4 \ \mathrm{TeV}^{-1}
\end{equation}
have been excluded over range $m_a = 5-100$ GeV, including the most
stringent constraints in the $5-10$ GeV region
\cite{CMS_LbL_axion:2025}. The current limits on the ALP
coupling to photons versus ALP mass $m_a$ are shown in
Fig.~\ref{fig:ALP_coupling_vs_mass} taken from
\cite{CMS_LbL_axion:2025}.

As one can see from Figs.~3 and 4, our best 95\% C.L. limit on the
ALP coupling for the ALP masses around 1 TeV is at least one order
of magnitude stronger than the current LHC limits. Thus, we can
conclude that the FCC-hh has a great physics potential of searching
for the heavy ALPs in proton-proton and ion-ion (proton-ion)
collisions.




\end{document}